\UseRawInputEncoding
\RequirePackage{fix-cm}
\documentclass[smallextended]{svjour3}       
\smartqed  
\usepackage{appendix}
\usepackage{amsmath}
\usepackage{graphicx}
\usepackage{lineno}
\usepackage{array}
\usepackage{longtable}
\usepackage{cite}

\journalname{Plasmonics}

\newcommand*\patchAmsMathEnvironmentForLineno[1]{%
\expandafter\let\csname old#1\expandafter\endcsname\csname #1\endcsname
\expandafter\let\csname oldend#1\expandafter\endcsname\csname end#1\endcsname
\renewenvironment{#1}%
{\linenomath\csname old#1\endcsname}%
{\csname oldend#1\endcsname\endlinenomath}}%
\newcommand*\patchBothAmsMathEnvironmentsForLineno[1]{%
\patchAmsMathEnvironmentForLineno{#1}%
\patchAmsMathEnvironmentForLineno{#1*}}%
\AtBeginDocument{%
\patchBothAmsMathEnvironmentsForLineno{equation}%
\patchBothAmsMathEnvironmentsForLineno{align}%
\patchBothAmsMathEnvironmentsForLineno{flalign}%
\patchBothAmsMathEnvironmentsForLineno{alignat}%
\patchBothAmsMathEnvironmentsForLineno{gather}%
\patchBothAmsMathEnvironmentsForLineno{multline}%
}

\begin{document}

\title{Decay dynamics of Localised Surface Plasmons: damping of coherences and populations of the oscillatory plasmon modes
}


\author{Krystyna Kolwas
}


\institute{\at Institute of Physics, \\Polish Academy of Sciences  \\
              Al. Lotnikow 32/46 \\
              02-668 Warsaw \\
              Tel.: +48 843 66 01 (ext. 3277)\\
              \email{Krystyna.Kolwas@ifpan.edu.pl}           
           }

\date{Received: DD Month YEAR / Accepted: DD Month YEAR}

\maketitle

\begin{abstract}

Properties of plasmonic materials are associated with surface plasmons - the electromagnetic excitations coupled to coherent electron charge density oscillations on a metal/dielectric interface.
Although decay of such oscillations cannot be avoided, there are prospects for controlling plasmon damping dynamics. In spherical metal nanoparticles (MNPs) the basic properties of Localized Surface Plasmons (LSPs) can be controlled with their radius.
The present paper handles the link between the size-dependent description of LSP properties derived from the dispersion relation based on Maxwell's equations and the quantum picture in which MNPs are treated as "quasi-particles".  Such picture, based on the reduced density-matrix of quantum open systems ruled by the master equation in the Lindblad form, enables to distinguish between damping processes of populations and coherences of multipolar plasmon oscillatory states and to establish the intrinsic relations between the rates of these processes, independently of the size of MNP.
The impact of the radiative and the nonradiative energy dissipation channels is discussed.

\end{abstract}

\section{Introduction}
\label{intro}

Plasmonics is a promising research area with many potential applications ranging from photonics, chemistry, medicine, bioscience, energy harvesting and communication to information processing
(e. g. \cite{barnes2003surface,xiong2007two,wang2011foundations,zhao2017cavity,rong2008resolving,xu2014nanoscale,lee2015aluminum,atwater2010plasmonics,couture2013modern,li2015plasmon,kolwas2017modification,mcphillips2010high,dykman2016biomedical,kolwas2016tailoring,scholl2012quantum}) and quantum optics
(e. g.  \cite{schirmer2004constraints,waks2010cavity,van2012spontaneous,martin2014large,delga2014quantum,doost2014resonant,torma2014strong,esteban2014strong,saez2017enhancing,hughes2018quantized,lobanov2018resonant}).
Plasmonics is based on the excitation of plasmons - electromagnetic excitations coupled to electron charge density oscillations on metal-dielectric interfaces, which result in confinement and enhancement of electromagnetic (EM) fields at metal/dielectric interfaces.

The optical features of metal nanoparticles (MNPs) are dominated by their ability to resonate with the EM radiation.
Excitation of collective surface charge oscillations in form of the standing waves known as Localized Surface Plasmons (LSPs) gives rise to a variety of effects such as the near-field concentration to the region well below the diffraction limit or the far-field resonant absorption and scattering in the spectral ranges which can be manipulated by MNP's dimensions, shape and the dielectric properties of the environment
\cite{bingham2010gas,anker2010biosensing,stiles2008surface,larsson2009nanoplasmonic,novo2008direct}.

Size-dependent spectral properties of noble MNPs (e.g. \cite{mulvaney1996surface,kreibig1995optical,hartland2011optical,kelly2003optical})
 and plasmon dynamics are of basic importance in many studied plasmonic issues.
The example can be harnessing hot electrons and holes, resulting from the decay of LSPs which recently  attracted considerable attention because of their promising applications in photodetection, energy-harvesting, hot-carrier-induced chemistry, photocatalysis, etc. \cite{clavero2014plasmon,brongersma2015plasmon,knight2011photodetection,brown2015nonradiative,gong2015materials,dal2018material,deeb2017plasmon,rajput2017investigation}. The performance of plasmonic devices correlate with numerous parameters that need to be studied to reach the optimum and desired properties. In all of them, LSP dynamics tailored by MNPs size play a crucial role.

The shapes and the spectral widths of the MNPs spectra are usually studied on the ground of the formalism known as Lorentz-Mie scattering theory with the solutions in the form of a sum of an infinite series of the spherical multipole partial waves.
Such fully classical electrodynamic description
of scattering of a plane wave  by a sphere
(e.g. \cite{mie1908,bohren1983absorption,born2005principles,quinten2010optical,hergert2012themie}) have been extended for the case of non-absorbing hosts
\cite{kolwas2017modification}, multi-layered spheres
\cite{bhandari1985scattering,sinzig1994scattering}, and arbitrary incident light beams \cite{gouesbet2011generalized}. LSP damping rates manifested in the broadening of the NPs spectra were experimentally studied in the case of the dominant dipole contribution to the spectra for NPs with available, limited sizes \cite{kreibig1995optical,
link1999spectral,heilweil1985nonlinear,sonnichsen2002plasmon,sonnichsen2002drastic,hartland2011optical}
(and references therein).
The resulting damping times and the frequencies corresponding to the maxima in the far-field intensity signals) are often suggested to directly characterize the LSP damping process and LSP resonance position.

However, the desired parameters (both for fundamental reasons and with a view to applications) such as size-dependent oscillation frequencies of multipolar modes corresponding to LSP's resonances,
or the decay rates (times) of the excited oscillations, are not the explicit parameters of Lorentz-Mie scattering theory.
In \cite{kolwas2009size,kolwas2010plasmonic,kolwas2013damping,derkachova2015dielectric}
such functions were derived from considering the classical dispersion relation \cite{ruppin1982electromagnetic,fuchs1992basic} for the surface localized EM (SLEM) fields basing on the self-consistent Maxwell divergent-free equations.
The modelling  \cite{kolwas2009size,kolwas2010plasmonic,kolwas2013damping,derkachova2015dielectric} provides the explicit size dependence of the oscillation frequencies of SLEM fields
and of the damping rates of such oscillations 
for the dipole and higher order multipolar surface modes and goes beyond the limitations related to the MNP's size including those which result from the often used quasistatic approximation  (e.g. \cite{kelly2003optical}) which kills the size dependence
(see also  \cite{waks2010cavity} in case of cavity QED formalism using the electrostatic electric field potential).

However, the such classical description of the dynamics of LSPs damping imposes constraints to fully understand the LSP dynamics including the damping processes, because it does not distinguish between the dephasing and population damping. In case of the dipole LSP such quantities were suggested to be connected to each other \cite{heilweil1985nonlinear,sonnichsen2002plasmon,sonnichsen2002drastic,hartland2011optical} and composed of radiative and nonradiative decay into electron-hole excitations. However, more detailed analysis of the problem was not given.

The present paper is aimed to present such an analysis by adopting the theory of quantum open systems and formulating the conclusions which apply to the issue of the  LSP damping.

Plasmonic phenomena are inherently quantum (see e.g. \cite{jacob2011plasmonics,tame2013quantum,bozhevolnyi2017plasmonics}).
Quantum plasmonics has recently emerged as a new fascinating field of research with a view of observing quantum phenomena in light-matter interactions at the nanoscale (\cite{de2012quantum,tame2013quantum,torma2014strong,ginzburg2016cavity,marquier2017revisiting} for reviews). In particular,
there have been a large number of theoretical studies of interactions between a (dipole) emitters and confined plasmonic structures in weak or also in strong-coupling regime using different approaches ('macroscopic' QED using Green's functions, quasi-normal decomposition, resonant-state-expansion) \cite{waks2010cavity,van2012spontaneous,sauvan2013theory,martin2014large,delga2014quantum,doost2014resonant,esteban2014strong,torma2014strong,muljarov2016resonant,saez2017enhancing,hughes2018quantized,lobanov2018resonant}.
The confinement of light field to scales far below that met in the case of the conventional optics by metalic nanostructures enables to describe the interactions between atomic systems and plasmonic structures in the formalism of cavity QED and  (see e.g.  \cite{waks2010cavity}) to explore the potential of such systems \cite{monroe2002quantum}  for developing future quantum  technologies like single-photon transistor or for carrying quantum information  (see \cite{tame2008single,ginzburg2016cavity,marquier2017revisiting} for reviews).

In this paper we describe the intrinsic dynamics of LSP which includes the relaxation pathways leading to the dephasing and population damping of plasmon oscillations within a quantum dynamical description of open systems. The master equation in the Lindblad form within the Born-Markov approximations \cite{breuer2002theory} is used to describe the evolution of an open quantum system of $N$ electrons confined in an MNP in absence of the driving EM field. Conclusions allow to generalize the common understanding of the LSP damping as the process which consists not only from the dephasing of LSP oscillations, but also from the population damping at the twice larger rate, irrespective the MNPs size and LSP multipolarity.

The results of such modeling have been related to the previously derived multipolar damping rates versus MNPs radius which describe the LSP decoherence process. Such rates resulted from the dispersion relation for the surface localized electromagnetic (SLEM) fields
\cite{kolwas2009size,kolwas2010plasmonic,kolwas2013damping,derkachova2015dielectric} which we shortly reconsider for completeness of the modelling. Derived in absence of the illuminating light, the dephasing and population damping rates tailor the transient LSP dynamics but manifest also in the amplitudes and widths of the spectra. The size dependence of studied parameters allows predicting the optimum and desired properties of LSPs, being a useful tool in tailoring MNPs plasmonic performance in experiments and applications also in the size regions still practically unavailable.

\section{Plasmon oscillatory eigenstates (classical description)} \label{sec:class}

Energy levels of atoms and molecules manifest in transitions from one energy
level to another when they absorb and emit light. We use a similar
picture of energy levels for a plasmonic system of $N$ free-electrons
confined in a spherical MNP.
\begin{figure}[htbp]
\centering
  \includegraphics[height=5cm]{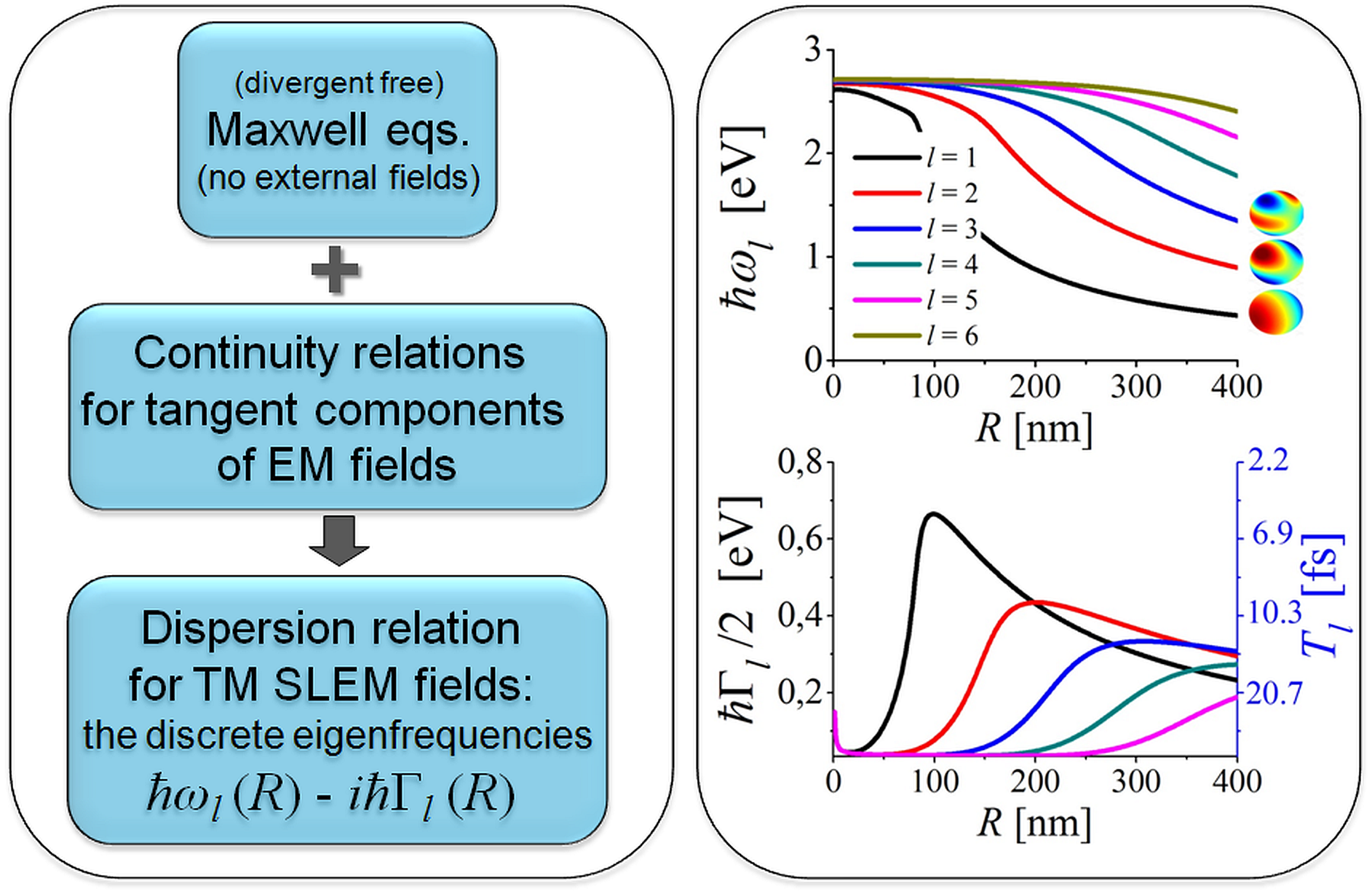}
  \caption{The scheme leading to the dispersion relation for SLEM fields (left) and the example of the resulting discrete complex eigenvalues $\hbar\omega_{l}-i\hbar\Gamma _{l}/2$ for the consecutive modes $l=1,2,3...$ (right) for Au MNPs embedded in water \cite{derkachova2015dielectric}. }
  \label{fig_eigenfrequencies}
\end{figure}
The starting point is based on the results of rigorous classical electrodynamic description based on the
self-consistent divergent free Maxwell equations (with no external sources). This problem was completely solved in a classical paper by Mie (1908) \cite{mie1908}.
However, in the present study based on the formulations
\cite{fuchs1992basic,boardman1977optical,kolwas2006smallest,derkachova2007size,kolwas2010plasmonic,kolwas2013damping,doost2014resonant,kolwas2009size,derkachova2015dielectric}, unlike in the case of the popular Lorentz-Mie scattering theory (e.g.
\cite{mie1908,bohren1983absorption,born2005principles,quinten2010optical,hergert2012themie}), the radiation illuminating the MNP is absent.
A spherical metal/dielectric interface forms a cavity \cite{doost2014resonant} which allows excitation of diverse modes of the SLEM fields in the form of the standing waves adjusted to the MNP's dimensions.
Such modes can be excited after MNP is illuminated by the light within the appropriate frequency range.
The continuity relation at the spherical interface leads to a set of separate dispersion relations for each multipole mode of the partial wave.
The solutions exist for TM polarized modes only, because contrary to TE modes, TM
(or $p$ polarized electric waves)
posses the non-zero normal to the surface (radial) components of the
electric field, which are able to couple with the surface free charges.

The dispersion relations for SLEM fields define the complex, discrete eigenvalues $\hbar({\omega }_{l}-i\Gamma _{l}/2)$ and connect the allowed oscillation frequency ${\omega }_{l}$ and the corresponding damping rate $\Gamma _{l}$ of the $l$'th mode oscillations to the (inverse
of) NS's radius $R$. The size dependence of $\omega_{l}(R)$ and $\Gamma_{l}/2(R)$
define the dynamics of SLEM fields: $\exp {(i\omega }_{l}-\Gamma _{l}/2)t)$ and is unambiguously determined by the material properties of the MNP and its dielectric environment.
Found in absence of the illuminating radiation, ${\omega }_{l}$ and $\Gamma _{l}$ inherently characterize an MNP of the radius $R$ in the same way as the energy levels and the inverse of lifetimes characterize an atom or a molecule. In both cases, these quantities manifest in the spectra, when the systems are illuminated. The example of ${\omega }_{l}(R)$ and $\Gamma _{l}(R)$ dependence
for gold MNPs embedded in water \cite{derkachova2015dielectric} is shown in Figure \ref{fig_eigenfrequencies}.


Let us note, that the problem of eigenmodes for plasmonic resonators recently attracted new interest resulting in a number of theoretical studies of quasinormal modes for plasmonic resonators and open cavities. In particular, based on the concept of the resonant-state expansion (e.g. \cite{doost2014resonant,zhang2016quantum,yan2018rigorous,dezfouli2018regularized,lobanov2018resonant}),  it was confirmed, that an open optical system like a plasmonic confined structure can be characterized by the complex
eigenfrequencies with the real and imaginary parts corresponding to, respectively, the spectral positions of the resonances and defining the spectral linewidths of resonances.  However, such a classical description does not include populations of the plasmon modes nor their damping.

\begin{figure*}[t]
 \centering
 \includegraphics[height=5.5cm]{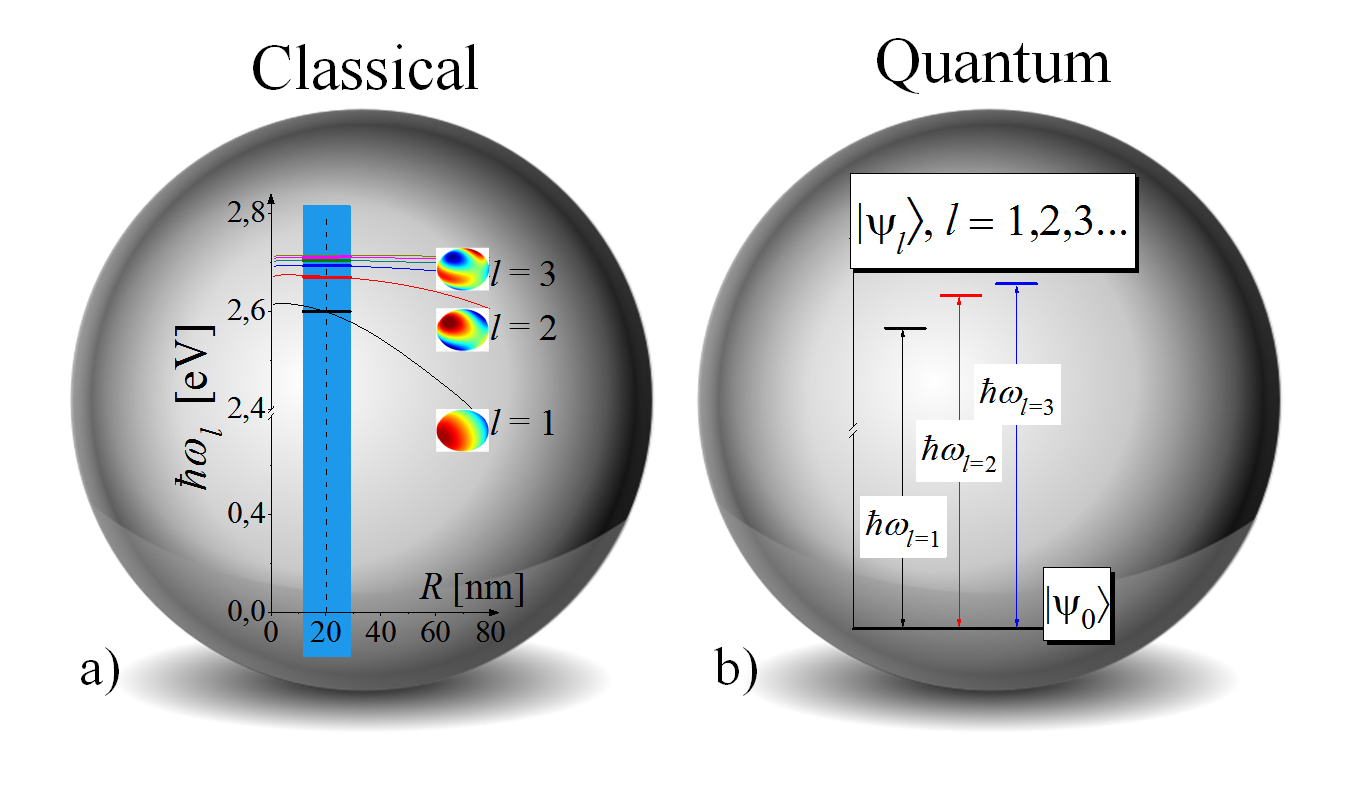}
 \caption{Ascribing the energy levels (Fig. b))  to the oscillation energies $\hbar\omega_{l}(R)$ (Fig. a)) which resulted from the dispersion relation for SLEM fields.}
 \label{fig_levels_class_quant}
\end{figure*}

The next step is to replace  the oscillation energies $\hbar\omega_{l}$ (or $\hbar\omega_{l}^r$) of the classical modes by the discrete energy levels, which are distinct from the zero-energy non-oscillatory level by the energies $\hbar {\omega }_{l}(R)$ (or $\hbar {\omega }_{l}^{r}(R)$)
(see Figure \ref{fig_levels_class_quant}a),b). The corresponding states of the plasmonic systems $S$ in the Hilbert space are
$|\psi _{l}\rangle $ with $l=1,2,3...$ (see Figure \ref{fig_levels_class_quant}c)). The only possible transitions are those with the absorption or emission of a photon with the energy
$\hbar {\omega }_{l}$.
Such transitions occur between the
state $|\psi _{l}\rangle $ and the non-oscillatory state $|\psi _{0}\rangle $.

\section{The density matrix for $N$ electrons of the plasmonic system and the quantum master equation} \label{DensityMatrix}

The density matrix (the density operator) is an alternate representation
of the state of a quantum system which is very convenient for systems in
the mixed states and in time-dependent problems. One of the reasons to
consider the mixed states is the entanglement of the systems with the
environment. The diagonal
elements of the density matrix correspond to the probabilities $%
p_{n}=N_{n}/N $ of occupying a quantum states $|\psi _{n}\rangle $, $n=$0,1,2... ,
so they describe the relative populations of these states.
The complex off-diagonal elements of the density matrix in the basis \{$|\psi _{l}\rangle ,|\psi _{0}\rangle $\} contain a time-dependent phase factors that describe the evolution of coherent superposition of the states.

The plasmonic system $S$ we consider consists of $N$ electrons
which in general can be distributed over the states $|\psi
_{n}\rangle $ , $n=0,l=1,2...$ with $|\psi_{0}\rangle$ for the ground, nonoscillatory states and $|\psi_{l}\rangle$ for the excited oscillatory states of the plasmon modes $l$ (see Figure \ref{fig_levels_class_quant}). Coherences between the states $|\psi_{0}\rangle $ and $|\psi _{l}\rangle $ can be created after the interaction of the system e.g.  with the external EM field.
As no physical system is absolutely isolated from its surroundings, the plasmonic system $S$ has to be considered as an open quantum system which is a subsystem of a larger combined quantum system $S+E$, where $E$ represents the environment to which the open system $S$ is coupled. Following the main assumption of the basic theory of open quantum systems \cite{breuer2002theory}, the environment is assumed to be a large system with an infinite number of degrees of freedom, (with a continuous and wide spectrum of characteristic frequencies) in thermal equilibrium. Therefore, the state of the system $E$ is practically unaffected by coupling to the system $S$. The interaction of the open system $S$ with the environment  causes an irreversible behavior of the open system $S$ and leads to decoherence (randomization of phases) and dissipation of energy into the surroundings.

The evolution of the open quantum system $S$
can be described by using a master equation which defines the dynamics of the reduced density operator $\rho^{S}(t)=tr_{E}\rho (t)$ obtained after tracing out the environment degrees of freedom (e.g. \cite{breuer2002theory}).
Quantum Markovian process represents the simplest case of the dynamics of an open systems. Within the Markov approximation, the memory effects of the system under influence of the environment are negligible. The correlation functions of the environment decay sufficiently fast over the time $\tau _{E}$ which is small compared to the characteristic timescale of the relaxation processes $\tau _{S}$ of the system $S$ (e.g. \cite%
{bertlmann2006open,breuer2002theory,blum2012density,li2018concepts}).

The dynamics of open systems in the case of Markov processes can be described by a first-order linear differential equation for the reduced density matrix , which is known as a quantum Markovian master equation in Lindblad form
\cite{lindblad1976generators,gorini1976completely,li2018concepts}:

\begin{equation}
\frac{\partial \rho ^{S}(t)}{\partial t}=-\frac{i}{\hbar }\left[ H,\rho
^{S}(t)\right] -D[\rho ^{S}(t)]  \label{Lindband}
\end{equation}

where $\rho ^{S}(t)$ is the reduced density matrix of the system $S$, and $D[\rho ^{S}(t)]$ is the so-called dissipator:

\begin{equation}
D[\rho ^{S}]=\frac{1}{2}\sum_{k}\left( L_{k}^{\dagger }L_{k}\rho ^{S}+\rho
^{S}L_{k}^{\dagger }L_{k}-2L_{k}\rho ^{S}L_{k}^{\dagger }\right)
\label{dissipator}
\end{equation}

Summation over $k$ extends  over all processes of coupling with the environment. The first term on the right-hand side of the eq. (\ref{Lindband}) describes the unitary evolution of the system $S$ under the action of a Hamiltonian $H$.
The dissipator $D[\rho ^{S}]$ describes the environmental influence on the system state (e.g. \cite%
{breuer2002theory}). The `jump' operators $L_{k}$ (eq. (\ref{dissipator})) describe a random evolution of the system which suddenly (at the time scale of the evolution) changes under the influence of the environment. Each $L_{k}\rho ^{S}L_{k}^{\dagger }$ term
induces one of the possible quantum jumps, while the remaining terms are needed to normalize properly the case in which no jump occurs.


The simplest quantum system is a two-level system whose Hilbert space is spanned by two states, an excited state and a ground state.
Such a two-level system is a very important basic model for an atom or a system of spins which is often used in quantum mechanics.
The two-level system can be successfully used, provided that the transitions to other levels can be neglected.

In case of LSPs such picture can be related to the classical description based on Maxwell equations (see Section \ref{sec:class}), where the problem is solved separately for each EM mode $l$ of the SLEM field, and the final result is a sum over the solutions for consecutive modes with $ l=1,2,3...$. Basing on such analogy, we describe the plasmonic system $S$ as a sum of $S_{l}$ of independent, open subsystems:
\begin{equation}
  S=\sum_{l=1}S_{l}
\end{equation}
Each subsystem $S_{l}$ is a two-level system:
the excited states $|\psi _{l}\rangle $ and the ground non-oscillatory state $|\psi_{0}\rangle$.
The state of each subsystem $S_{l}$ can be described by the 2$\times $2 matrix operator $\rho ^{S_{l}}(t)$.
Each system $S_{l}$ is coupled to the environment
independently and all assumptions about the coupling of the system $S_l$ to the environment remain fulfilled for subsystems $S$.
The dynamics of the system $S$ thus results from the independent dynamics of the systems $S_{l}$:

\begin{equation}
\rho ^{S}(t)=\sum_{l}\rho ^{S_{l}}(t)  \label{ro_S}
\end{equation} \label{ro_sum}

governed by the Lindblad eq. (\ref{Lindband}). The form of the Lindblad equation guarantees, that also dynamics
of each matrix operator $\rho ^{S_{l}}(t)$ is governed by the equation in the Lindblad form.
Therefore we can use the standard basis \{$|\psi _{l}\rangle ,|\psi _{0}\rangle $\} represented by the
two-dimensional column vectors:

\begin{equation}
|\psi _{0}\rangle =\left(
\begin{array}{c}
0 \\
1%
\end{array}%
\right) ,\ |\psi _{l}\rangle =\left(
\begin{array}{c}
1 \\
0%
\end{array}%
\right)   \label{vector_basis}
\end{equation}%
and represent the operator $\rho ^{S_l}(t)$ in matrix form:

\begin{equation}
\rho ^{S_{l}}(t)=\left(
\begin{array}{cc}
\rho _{ll}(t) & \rho _{l0}(t) \\
\rho _{0l}(t) & \rho _{00}(t)%
\end{array}%
\right)   \label{ro_S_l}.
\end{equation}

The diagonal elements of the matrix $\rho ^{S_l}$ represent the relative populations (more precisely, the population probability densities) of the relevant states. The off-diagonal elements represent quantum-mechanical coherences and are complex conjugates of each other, carrying the same information.

The Hamilton operator $H_{l}\equiv E_{l}|\psi _{l}\rangle \langle \psi
_{l}|-E_{0}|\psi _{0}\rangle \langle \psi _{0}|)$ with the  energy eigenvalue $E_{0}=0$ (Figure \ref{fig_levels_class_quant}) in the ground state ($H_{l}|\psi _{0}\rangle =E_{0}|\psi _{0}\rangle=0$):

\begin{equation}
H_{l}|\psi _{l}\rangle =E_{l}|\psi_ {l}\rangle
\end{equation}

in the chosen basis is represented by the matrix:

\begin{equation}
H_{l}=E_{l}\left(
\begin{array}{cc}
1 & 0 \\
0 & 0%
\end{array}%
\right) .  \label{H_l}
\end{equation}

The commutator $\left[ H_{l},\rho ^{S_{l}}\right] $ possess the non-zero off-diagonal elements only.

The transition operator from the state
$|\psi _{l}\rangle$
to the state
$|\psi _{0}\rangle $: $\sigma _{-}|\psi
_{l}\rangle =|\psi _{0}\rangle \langle \psi _{l}||\psi _{l}\rangle =|\psi
_{0}\rangle $,
and its complex conjugate $\left( \sigma _{-}\right)
^{\dagger }=\sigma _{+}$ ($\sigma _{+}|\psi _{0}\rangle =|\psi _{l}\rangle
\langle \psi _{0}|\psi _{0}\rangle =|\psi _{l}\rangle $)
can be expressed in the basis
\{$|\psi _{l}\rangle ,|\psi _{0}\rangle $\}
with the use of the Pauli $2\times 2$ matrices
$\sigma _{1}$ and $\sigma _{2}$:

\begin{equation}
 \sigma _{-}=\frac{1}{2}\left( \sigma _{1}-i\sigma _{2}\right) =\left(
\begin{array}{cc}
0 & 0 \\
1 & 0%
\end{array}%
\right) ,\text{  \ }\sigma _{+}=\frac{1}{2}\left( \sigma _{1}+i\sigma
_{2}\right) =\left(
\begin{array}{cc}
0 & 1 \\
0 & 0%
\end{array}%
\right) \label{Pauli}
\end{equation}

which are eigenoperators of
the Hamiltonian $H_{l}$ (eq. (\ref{H_l})), which describe the emission and
absorption processes. As usual, $\sigma _{-}$ is the operator lowering the energy,
$\sigma _{+}$ is the energy rising operator: $\left[
H_{l},\sigma _{\mp }\right] =\mp \hbar \omega _{l}\sigma _{\mp }$.

\subsection{Dynamics of the radiative plasmon damping}

An excited plasmon, similarly as an excited atom, decays to the state of lower energy spontaneously emitting a photon. In the theory of open quantum systems, such decay is assumed to be due to the coupling of the system to the EM vacuum (fluctuating) fields. The random vacuum fluctuations (which pervades all space even at zero temperature) cause random jumps influencing plasmon evolution.
In the case when only radiative damping is present, the Hamiltonian in the Linblad eq. (\ref{Lindband}):
$H_{l}^{r}=H_{l}$ (eq. (\ref{H_l})) and the eigenenergy $E_{l}=\hbar \omega_{l}^{r}$.

A single jump operator
\begin{equation}
L_{l}^{r}=\sqrt{\Gamma _{l}^{r}}\sigma _{-}
\end{equation}
describs random, sudden emission of a photon from the state $|\psi _{l}\rangle $  to the state $|\psi _{0}\rangle $ of the $S_{l}$, with
$\Gamma _{l}^{r}$ being the (spontaneous) radiation rates.
The Lindblad master equation (\ref{Lindband}) takes then the form:

\begin{equation}
\frac{\partial \rho ^{S_{l}}}{\partial t}=-\frac{i}{\hbar }\left[
H_{l}^{r},\rho ^{S_{l}}\right] -\frac{1}{2}\Gamma _{l}^{r}\left( \sigma
_{+}\sigma _{-}\rho ^{S_{l}}+\rho ^{S_{l}}\sigma _{+}\sigma _{-}-2\sigma
_{-}\rho ^{S_{l}}\sigma _{+}\right)  \label{masterS_l_rad}.
\end{equation}

The evolution of the diagonal and off-diagonal elements of $\rho ^{S_{l}}$

\begin{equation}
\frac{d}{dt}\left(
\begin{array}{cc}
\rho _{ll} & \rho _{l0} \\
\rho _{0l} & \rho _{00}%
\end{array}%
\right) =i\omega _{l}^{r}\left(
\begin{array}{cc}
0 & \rho _{l0} \\
-\rho _{0l} & 0%
\end{array}%
\right) -\Gamma _{l}^{r}\left(
\begin{array}{cc}
\rho _{ll} & \frac{1}{2}\rho _{l0} \\
\frac{1}{2}\rho _{0l} & -\rho _{ll}%
\end{array}%
\right)
\end{equation}
leads to the solutions ($t_0 = 0$):

\begin{eqnarray}
\rho _{ll}(t) &=&\rho _{ll}(t_0)\exp \left( -\Gamma _{l}^{r}t\right), \label{pop}\\
\rho _{00}(t) &=&\rho _{ll}(t_0)\left( 1-\exp \left( -\Gamma _{l}t\right)\right), \\
\rho _{0l}(t) &=&\rho _{0l}(t_0) \exp
\left( -i\omega _{l}^{r}-\Gamma _{l}^{r}/2\right) ,\\
\rho _{l0}(t) &=&\rho _{l0}(t_0)\exp \left( i\omega _{l}^{r}-\Gamma_{l}^{r}/2\right)t. \label{coh}
\end{eqnarray}

Therefore, the radiative damping rate $\Gamma _{l}^{r\thinspace pop}=\Gamma _{l}^{r}$ of populations $\rho _{ll}$ is twice as large as the radiative rate $\Gamma _{l}^{r\thinspace coh}=\Gamma _{l}^{r}/2$ of coherences $\rho _{l0}=\rho _{0l}^*$, so the relation between the corresponding radiative lifetimes of populations and coherences are $2T_{l}^{r\thinspace pop}=$ $T_{l}^{r\thinspace coh}$.

\subsection{Dynamics of the total (radiative and collisional) plasmon damping}

Electrons in a metal inevitably undergo collisions which lead to the dissipation of energy and release of heat.
To account for nonradiative relaxation processes, the system $S$ interacting with fluctuations of the vacuum is assumed to be immersed in a dissipative heat-bath in a thermal equilibrium state with an infinite number of degrees of freedom. The heat reservoir dynamics is assumed to be much faster than those of the open system $S_{l}$, so the dynamics of $S_{l}$ (and those of $S$) is Markovian. The jump operators which describe the collisional transition from the state $|\psi _{l}\rangle $ to the state $|\psi_{0}\rangle $ are:

\begin{equation}
L_{l}^{n}=\sqrt{\Gamma _{l}^{nr}}\sigma _{-}
\end{equation}

where $\Gamma _{l}^{nr}$ are the nonradiative rates describing the collisional processes leading to the heat release.
Summing over radiative and nonradiative contributions in the dissipator $D[\rho ^{S}]$ (eq. (\ref
{dissipator})) we get the master equation:

\begin{equation}
\frac{\partial \rho ^{S_{l}}}{\partial t}=-\frac{i}{\hbar }\left[ H_{l},\rho
^{S_{l}}\right] -\frac{1}{2}\left( \Gamma _{l}^{r}+\Gamma _{l}^{nr}\right)
\left( \sigma _{+}\sigma _{-}\rho ^{S_{l}}+\rho ^{S_{l}}\sigma _{+}\sigma
_{-}-2\sigma _{-}\rho ^{S_{l}}\sigma _{+}\right)  \label{masterS_l}
\end{equation}

The random jumps in the evolution of $\rho ^{S_{l}}(t)$ with the rates $\Gamma _{l}^{r}$ and $\Gamma _{l}^{nr}$ are assumed to be uncorrelated. The relaxation of the diagonal and off-diagonal elements of $\rho ^{S_{l}}$ is governed by the total relaxation rate $\Gamma _{l}=\Gamma_{l}^{r}+\Gamma _{l}^{nr}$.
The Hamiltonian  $H_{l}$ (eq. (\ref{H_l})) defines the eigenenergies $E_{l}=\hbar \omega _{l}$
of the excited states $|\psi _{l}\rangle $ in presence of the radiative and
nonradiative damping processes.

Using the formalism recalled in the previous section and the equation \ref{ro_S}, the evolution of the whole system $S$ is described by the dynamics of the
density matrix $\rho ^{S}(t)$, (eqs. (\ref{ro_S}, \ref{ro_S_l})):
\begin{equation}
    \rho ^{S}(t)=
    \left(
\begin{array}{cc}
\sum\limits_{l}\rho _{ll}(t) &\sum\limits_{l} \rho _{l0}(t) \\
\sum\limits_{l}\rho _{0l}(t) & \rho _{00}(t)%
\end{array}%
\right)  \label{ro_S_t}
\end{equation}

with the diagonal and off-diagonal temporal dependence given by:

\begin{equation}
\rho ^{S}(t)=\left(
\begin{array}{cc}
\sum\limits_{l}\rho _{ll}(t_{0})\exp \left( -\Gamma _{l}t\right)  &
\sum\limits_{l}\rho _{l0}(t_{0})\exp \left( i\omega _{l}t-\frac{1}{2}\Gamma
_{l}t\right)  \\
\sum\limits_{l}\rho _{0l}(t_{0})\exp \left( -\left( i\omega _{l}t-\frac{1}{2}\Gamma _{l}t\right) \right)  & \sum\limits_{l}\rho _{ll}(t_{0})\left(
1-\exp \left( -\Gamma _{l}t\right) \right)
\end{array}%
\right). \label{ro_S(t)}
\end{equation}

The populations of the oscillatory states are exponentially damped with the rates:

\begin{equation}
\Gamma _{l}^{pop}=\Gamma _{l}^{r}+\Gamma _{l}^{nr},
\end{equation}
while the oscillations of coherences are damped
with two times lower rate:
\begin{equation}
\Gamma _{l}^{coh}=\frac{1}{2}\Gamma _{l}^{pop}=\frac{1}{2}\left( \Gamma
_{l}^{r}+\Gamma _{l}^{nr}\right) .  \label{Gamma_coh_pop}
\end{equation}

So the lifetime of coherences is twice as large as the lifetime of populations.

The temporal evolution of coherences $\rho ^{S_{l}}$ proceeds according to $ \exp (i\omega _{l}-\Gamma _{l}/2)t$, similarly to the evolution of SLEM fields, which resulted from the dispersion relation (Section \ref{sec:class}).
So, the size dependence of $\omega_l$ and $\Gamma_l$ in eq. (\ref{ro_S(t)}) can be found by solving the dispersion relation for SLEM field for successive $R+\Delta R$, as it was done in our previous papers \cite{kolwas2009size,kolwas2010plasmonic,kolwas2013damping,derkachova2015dielectric}.

The density matrix $\rho ^{S}(t)$ (eq. (\ref{ro_S(t)})) fulfills all the properties which are of basic importance for quantum statistics: it is positive, self-adjoint and with the trace equal 1.
The relative populations of the states $|\psi_{n}\rangle $,  $n=0,l$:
\begin{equation}
    \sum\limits_{l}N_n(t) /N = \rho_{00}(t) +\sum\limits_{l}\rho_{ll}(t)
\end{equation}

are the same at any time $t$ as the initial populations
\begin{equation}
    \sum\limits_{l}N_l(_0) /N =\sum\limits_{l}\rho_{ll}(t_0)=1
\end{equation}

over the \textit{excited}, oscillatory states $|\psi_{l}\rangle $, $l=$1,2,3... at $t=t_0$ (see eq. (\ref{ro_S(t)})):

\begin{equation}
Tr\left[ \rho ^{S}(t)\right] =
\rho_{00}(t) +\sum\limits_{l}\rho_{ll}(t)
=\sum\limits_{l}\rho
_{ll}(t_{0})=1.
\end{equation}

\subsection{Pure dephasing}

Pure dephasing takes place where the (external) reservoir is the source of fluctuations which do not change the average energy of the system. These fluctuations lead to a loss of coherence in the system resulting in the decay of the off-diagonal elements of the density matrix, without affecting the diagonal elements.
So, if there are processes which lead to decoherence without affecting populations, the total damping rate is increased by the "pure" decoherence rate $\Gamma _{l}^{\ast coh}$:

\begin{equation}
\Gamma _{l}^{coh}=\frac{1}{2}\Gamma _{l}^{pop}+\Gamma _{l}^{\ast coh}=\frac{1%
}{2}\left( \Gamma _{l}^{r}+\Gamma _{l}^{nr}\right) +\Gamma _{l}^{\ast coh}.
\end{equation}

 The population damping rates $\Gamma _{l}^{pop}$ remain unchanged:

\begin{equation}
\Gamma _{l}^{pop}=\Gamma _{l}^{r}+\Gamma _{l}^{nr}.
\end{equation}

In the case when the nanoparticle is illuminated by light, such 'pure'  dephasing processes introduce additional broadening to the observed spectra in which LSP excitations manifest. 

\section{Discussion and conclusions}

Localized Surface Plasmons in plasmonics are commonly described as the electromagnetic excitations coupled to coherent electron charge densities oscillations on a metal/dielectric interface.
LSP's parameters such as damping rates of plasmon oscillations in function of MNP's size (related to the decoherence processes) are usually derived from the widths of scattering or absorption spectra for the dipole ($l=$1) plasmon mode only.   In common practice, such damping rates (and the frequencies corresponding to the maxima in the far-field intensity signals) are suggested to describe LSP dynamics what may lead to narrow understanding of LSP damping, which in fact is the process embracing dephasing and depopulation of all the plasmon modes involved, including those not manifested as distinct maxima in the spectra. Moreover, it is known that  in general LSPs resonances can manifest in different manner in divers spectra in the near- and far-field regions \cite{zuloaga2011energy,kats2011effect,alonso2013experimental,moreno2013analysis,cacciola2016spectral}.

Solutions of the dispersion relation for SLEM fields (see the example in Figure \ref{fig_eigenfrequencies}) allow to find the intrinsic LSP dephasing rates and resonance frequencies of multipolar plasmons in absence of illumination and predict their size dependence in the large range of MNP radii \cite{kolwas2009size,kolwas2010plasmonic,kolwas2013damping,derkachova2015dielectric}.

In the applied quantum description an MNP is treated as "quasi-particle", similarly to an atom or molecule. Such picture enables studying the intrinsic decay dynamics of both: populations and coherences of the quasi-particles oscillatory states involving physical quantities and their relations which have not been discussed in the previous approaches. In particular,
the damping rates of populations and of coherences of consecutive plasmon modes occur to be intrinsically different: $\Gamma _{l}^{pop}=2\Gamma _{l}^{coh}$, regardless of the MNP size.

\begin{figure}[h]
\centering
  \includegraphics[height=5.5
  cm]{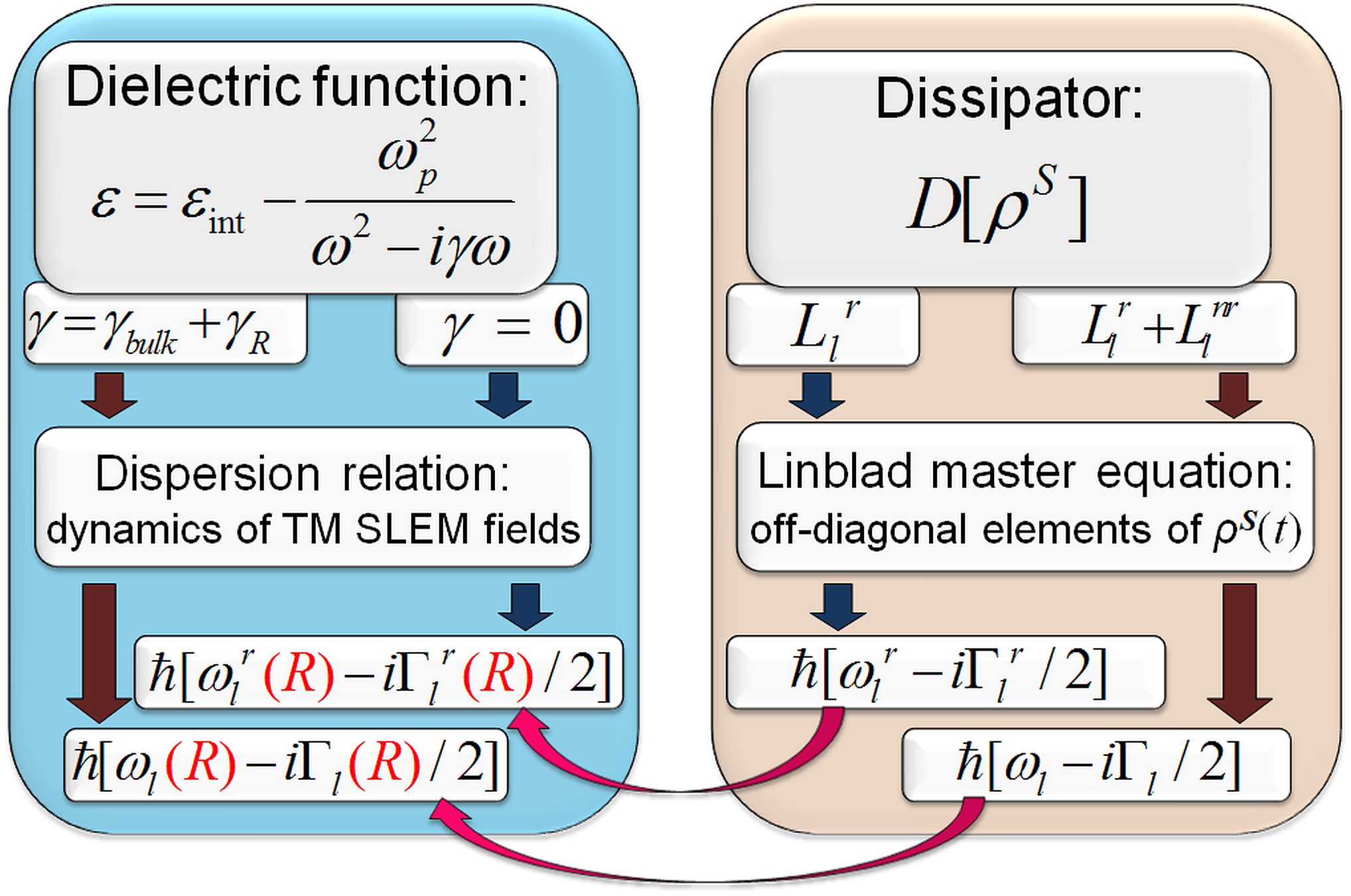}
  \caption{The correspondence of deriving the radius dependent quantities which describe the dynamics of the SLEM fields within classical EM modelling (left) and the corresponding scheme for the dynamics of coherences in the quantum description (right).}
  \label{fig_analogy}
\end{figure}
The present paper builds the bridge between the results of the classical electrodynamic description supplying size dependence of the dephasing rates and the quantum description of the intrinsic dynamics of plasmon decay processes in MNPs which are essentially not restricted in size (see Figure \ref{fig_analogy}).
Such dynamics is affected
by several factors to consider such as electron dumping in bulk metal $\gamma_{bulk}$, the electron-surface scattering $\gamma_R$ (the parameters of the dielectric function), radiation damping and prospectively the interface damping resulting in energetic charge carriers production.
In case of solving the dispersion relation for SLEM fields using the Drude-like dielectric function (electron dumping in bulk metal with the rate $\gamma_{bulk}$ and the radius dependent electron-surface scattering rate $\gamma_R$ accounted), the damping rates $\Gamma_l(R)$ define the size dependence of the total population and and coherence damping rates resulting from the quantum modelling: $\Gamma_l(R)=\Gamma_l^{coh}(R)=\Gamma_l^{pop}(R)/2$ (see Figure \ref{fig_analogy}).
Such link can be
a useful tool in tailoring MNPs plasmonic performance in experiments and applications also in the size regions still practically unavailable. Suggested (Figure \ref{fig_analogy}) decomposition of the contribution of the radiative rates from the total damping rates after appropriate modification of the input dielectric function of gold and silver ($\gamma_{bulk}=$0) will be the subject of our next study.

The quantum picture offers the attractive prospects for designing the plasmonic devices that operate at the quantum level and exploit their lossy nature (e.g. \cite{verstraete2009quantum,linic2011plasmonic,clavero2014plasmon,khurgin2015deal}), in addition to the broad range of applications based on the enhancement of EM fields at metal/dielectric interfaces.
In such a practical context, accounting for various dissipative decay channels and quantification of their importance in the plasmonic systems built of diverse material, size, and shape, embedded in various matrices, is of basic importance.

\newpage

\bibliographystyle{unsrt}

\bibliography{Density_matrix}

\end{document}